\documentstyle[12pt,openbib]{article}
\title{Black Hole Entropy is Noether Charge}
\author{Robert M. Wald\\
         {\it University of Chicago}\\
         {\it Enrico Fermi Institute and Department of Physics}\\
         {\it 5640 S. Ellis Avenue}\\
         {\it Chicago, Illinois 60637-1433}}
\date{}

\begin{document}
\maketitle
\begin{abstract}
We consider a general, classical theory of gravity in $n$ dimensions,
arising from a diffeomorphism invariant Lagrangian. In any such
theory, to each vector field, $\xi^a$, on spacetime one can associate
a local symmetry and, hence, a
Noether current $(n-1)$-form, ${\bf j}$, and
(for solutions to the field equations) a
Noether charge $(n-2)$-form, ${\bf Q}$, both of which are locally
constructed from $\xi^a$ and
the the fields appearing in the Lagrangian. Assuming only that the
theory admits stationary black hole solutions with a bifurcate Killing
horizon (with bifurcation surface $\Sigma$), and that the canonical mass
and angular momentum of solutions
are well defined at infinity, we show that
the first law of black hole mechanics always holds for perturbations to
nearby stationary black hole solutions. The quantity playing the role of
black hole entropy in this formula is simply $2 \pi$ times the integral
over $\Sigma$ of the Noether charge $(n-2)$-form
associated with the horizon Killing field (i.e., the Killing field
which vanishes on $\Sigma$), normalized so as to have unit surface
gravity. Furthermore, we show that this black hole entropy
always is given by a local
geometrical expression on the horizon of the black hole. We thereby
obtain a natural candidate for the entropy of a dynamical black hole
in a general theory of gravity.
Our results show
that the validity of the ``second law" of black hole mechanics in dynamical
evolution from an initially stationary black hole to a final stationary
state is equivalent to the positivity of a total Noether flux, and thus may
be intimately related to the positive energy properties of the theory.
The relationship between the derivation of our formula for black hole
entropy and the derivation via
``Euclidean methods" also is explained.

{\bf PACS \#: } 04.20.-q, 97.60.Lf
\end{abstract}
\newpage

	One of the most remarkable developments in the theory of black
holes in classical general relativity was the discovery of a close
mathematical analogy between certain laws of ``black hole mechanics"
and the ordinary laws of thermodynamics. When the effects of quantum
particle creation by black holes \cite{haw}
were taken into account, this analogy
was seen to be of a physical nature, and it has given rise to some deep
insights into phenomena which may be expected to occur in a quantum
theory of gravity.

	The original derivation of the laws of black hole mechanics in
classical general relativity \cite{bch} used many detailed properties of
the Einstein field equations, and, thus, appeared to be very special
to general relativity. However, recently it has become clear that at least
some of the laws of classical black hole mechanics hold in a much more
general context. In particular, it has been shown that a version of the
first law of black hole mechanics holds in any theory of gravity derivable
from a Hamiltonian \cite{sw}. (For the cases of $(1+1)$-dimensional
theories of gravity \cite{fro}
and Lovelock gravity \cite{jm}, the explicit forms of this law
have been given.) Furthermore, analogs of all of the
classical laws of black hole mechanics have been shown to hold in
$(1+1)$-dimensional theories \cite{fro}.

	However, despite the very general nature
of the Hamiltonian derivation
\cite {sw} of the first law of black hole mechanics, there remains
one unsatisfactory aspect of the status of the
first law in a general theory of gravity \cite{jac}: Although
the derivation shows that for a perturbation of a stationary
black hole, a surface integral at the black hole horizon (involving the
unperturbed metric and its variation) is equal to terms involving the
variation of mass and angular momentum (and possibly
other asymptotic quantities)
at infinity, the derivation does not show that this surface term at
the horizon can be expressed as
$\kappa / 2 \pi$ (where $\kappa$ denotes the unperturbed
surface gravity) times the variation of a
surface integral of the form $S = \int_{\Sigma} F$, where $F$ is
locally constructed out of the metric and other dynamical
fields appearing in the
theory. It is necessary that the horizon
surface term be expressible in this form in order to be able to
identify a local, geometrical
quantity, $S$, as playing the role of the entropy of the black hole.

	The main purpose of this paper is to remedy this deficiency by
showing that in a general theory of gravity derivable from a Lagrangian,
the form of the first law of black hole mechanics for perturbations to
nearby stationary black holes is such that the surface term at the horizon
always takes the form $\frac{\kappa}{2 \pi} \delta S$, where $S$ is a
local geometrical quantity, and is equal to $2 \pi$ times
the Noether charge at the horizon of the horizon Killing field (normalized
so as to have unit surface gravity).
The local, geometrical character of $S$ suggests a possible
generalization of the definition of entropy to dynamical black holes.
The relationship between black hole entropy and Noether charge also
suggests the possibility of a general relationship between the validity
of the second law of black hole mechanics (i.e., increase of black hole
entropy) and positive energy properties of a theory. An additional
byproduct of our analysis is that it will enable us to make contact with the
``Euclidean derivation" of formulas for black hole entropy, thereby
demonstrating equivalence of that approach with other
approaches -- a fact that is not at all easy to see by a direct comparison
of, say, references \cite{sw} and \cite{gh}. Our considerations
in this paper will be limited
to a general analysis of all
the above issues; applications to particular theories will be given
elsewhere \cite{iw}

	Before presenting our new derivation of the first law, we comment
upon the status of other ``preliminary laws" of black hole mechanics in a
general theory of gravity. We consider theories defined on an
$n$-dimensional manifold $M$ with dynamical fields consisting of a
(Lorentzian) spacetime metric, $g_{ab}$, and possibly other matter
fields, such that the equations of motion of the metric and other
fields are derivable from a diffeomorphism
invariant Lagrangian. (Our precise assumptions concerning the
Lagrangian will be spelled out in more detail below.) We assume that
a suitable notion of ``asymptotic flatness" is defined in the theory. The
{\it black hole} region of an asymptotically flat spacetime
then is defined to be the
complement of the past of the asymptotic region. In order to begin
consideration of the classical laws of black hole mechanics, it is
necessary that the event horizon of a stationary black hole be a Killing
horizon, i.e., a null surface to which a Killing vector
field is normal. This property is known
to be true in general relativity by a nontrivial argument using the null
initial value formulation \cite{he}, so it is not obvious that it would
hold in more general theories of gravity.
Nevertheless, this property automatically holds for all static
black holes (since the static Killing field must be normal to the event
horizon of the black hole), and, hence, it automatically holds for
spherically symmetric black holes (in the $O(n-1)$ sense)
and, thus, in particular, for all black holes
in $(1+1)$-dimensional theories of gravity.

	The {\it surface gravity}, $\kappa$, at any point, $p$, of a Killing
horizon ${\cal H}$ is defined by (see, e.g., \cite{wa}),
\begin{equation}
\xi^a \nabla_a \xi^b = \kappa \xi^b
\label{kappa}
\end{equation}
where $\xi^a$ is the Killing field normal to ${\cal H}$. The zeroth law of
black hole mechanics asserts that $\kappa$ is constant over the event
horizon of a stationary black hole. The proof of this law {\it does} make
direct use of the specific form of the Einstein field equations \cite{bch},
and, thus, does not appear likely to generalize to other theories of
gravity \cite{zl}.
Nevertheless, the zeroth law trivially holds for spherically symmetric
black holes, and, in particular, in all $(1+1)$-dimensional theories.

	It is worth noting that the validity of the zeroth law is, in essence,
equivalent to the statement that -- apart from the ``degenerate case" of
vanishing surface gravity -- the event horizon of a stationary black hole
must be of bifurcate type. Namely, it is easily proven that a bifurcate
Killing horizon must have constant (and nonvanishing) surface gravity,
whereas it can be shown \cite{rw} that any Killing horizon with constant,
nonvanishing surface gravity can be locally extended (if necessary)
to a bifurcate horizon.

	It also should be noted that in an arbitrary theory of gravity, a
black hole with constant surface gravity will ``Hawking radiate" at
temperature $\kappa / 2 \pi$ when quantum particle creation effects
are taken into account, i.e., the Einstein field equations play no role in
the derivation of the Hawking effect. Similarly, the theorems of \cite{kw}
on the uniqueness and thermal properties of quantum states
on black holes with bifurcate horizons hold in an arbitrary theory of
gravity. Thus, $\kappa / 2 \pi$
always represents the physical temperature of a black hole.

	We turn, now, to the presentation of a new derivation of the first
law of black hole mechanics for stationary black holes with bifurcate
horizons in a general theory of gravity
in $n$-dimensions derived from a diffeomorphism
invariant Lagrangian. We shall follow closely the framework of Lagrangian
field theories developed in \cite{lw}, with one small change: We shall view
the Lagrangian as an $n$-form, ${\bf L}$, rather than as a scalar
density; similarly, other tensor densities of \cite{lw} will appear here
in their dualized version as differential forms. In order to define
${\bf L}$, it is necessary to introduce a fixed (i.e., ``nondynamical")
derivative operator, $\nabla_a$, on spacetime. It also may be necessary
to introduce other ``non-dynamical, background fields",
$\gamma$, such as the curvature of $\nabla_a$ (if $\nabla_a$ is non-flat);
we shall assume, however, that any such additional fields,
$\gamma$, are uniquely
determined by $\nabla_a$, and that $\gamma$ changes by a diffeomorphism
under the change induced in $\nabla_a$
by the action of that diffeomorphism.
At each point
$p$ of spacetime, ${\bf L}$ then is
required to be a function of the spacetime metric, $g_{ab}$, (or,
alternatively, of a tetrad or soldering form)
and finitely many of its (symmetrized)
derivatives at $p$, as well as of other matter
fields present in the theory and their (symmetrized)
derivatives at $p$, and of $\gamma$ at $p$. Note that no
restriction is placed upon the number of derivatives of the metric or other
fields upon
which ${\bf L}$ can depend (other than that this number be finite),
so ``higher derivative" gravity theories are included in this framework.

	In order to
reduce the number of symbols and indices appearing in formulas, I
shall use the symbol ``$\phi$" to denote all of the dynamical fields,
including the spacetime metric. We shall restrict attention to
diffeomorphism invariant theories, by which
we mean that for any diffeomorphism, $\psi : M \rightarrow M$,
we have,
\begin{equation}
{\bf L}[\psi^* (\phi)] = \psi^* {\bf L}[\phi]
\label{dif}
\end{equation}
Note that on the left side of this equation, $\psi^*$ is
{\it not} applied to $\nabla_a$ or any other non-dynamical fields
$\gamma$ which may appear in ${\bf L}$. Equation (\ref{dif}) can
be interpreted as stating that -- although it may be necessary to
introduce $\nabla_a$ and/or $\gamma$ to define ${\bf L}$ --
${\bf L}$ actually depends only upon the dynamical fields $\phi$.

	 Under a first order variation of the dynamical fields, the variation
of ${\bf L}$ can be put in the form (see, e.g., \cite{lw}),
\begin{equation}
\delta {\bf L} = {\bf  E} \delta \phi + d {\bf \Theta}
\label{lag}
\end{equation}
where summation over the dynamical fields (and contraction of their
tensor indices with corresponding dual tensor indices of ${\bf E}$)
is understood in the first term on the right side of this equation.
The $(n-1)$-form, ${\bf \Theta}$,
is locally constructed from $\phi$
and $\delta \phi$, but is determined by eq.(\ref{lag}) only up
to addition of a closed (and, hence, exact \cite{w}) form locally constructed
from the fields appearing in ${\bf L}$; we shall adopt
eq. (2.12) of \cite{lw} as our definition of ${\bf \Theta}$.
The symplectic current $(n-1)$-form, ${\bf \Omega}$, is defined
in terms of the variation of ${\bf \Theta}$ by,
\begin{equation}
{\bf \Omega}(\phi, \delta_1 \phi, \delta_2 \phi) =
\delta_1 [{\bf \Theta}(\phi, \delta_2 \phi)] -
\delta_2 [{\bf \Theta}(\phi, \delta_1 \phi)]
\label{sym}
\end{equation}
It should be noted that ${\bf \Theta}$ and ${\bf \Omega}$ will depend
upon the choice of $\nabla_a$ in sufficiently high derivative theories
\cite{lw} -- although they change only by an exact form, i.e., a ``surface
term", under a change of derivative operator -- and they
need not be diffeomorphism invariant in the sense of eq.(\ref{dif}).
Furthermore, ${\bf \Theta}$ and ${\bf \Omega}$ will change
when an exact form is added to ${\bf L}$ -- with the change in
${\bf \Omega}$ being given by an exact form -- even though
such a modification of ${\bf L}$ has no effect upon the equations
of motion, ${\bf  E} = 0$.

	Now, let $\xi^a$ be any vector field on $M$ and consider the field
variation $\hat{\delta} \phi = {\cal L}_{\xi} \phi$.
The diffeomorphism invariance of
${\bf L}$ implies that under this variation, we have,
\begin{equation}
\hat{\delta} {\bf L} = {\cal L}_{\xi} {\bf L} = d (\xi \cdot {\bf L})
\label{ls}
\end{equation}
where here and below, we make frequent use of the general identity
\begin{equation}
{\cal L}_{\xi}
{\bf \Lambda} = \xi \cdot d{\bf \Lambda}  + d (\xi \cdot {\bf \Lambda})
\label{id}
\end{equation}
holding for any differential form ${\bf \Lambda}$ and vector field
$\xi^a$, where ``$\cdot$" denotes the contraction of a
vector field with the first index
of a differential form. Equation (\ref{ls}) shows that the vector fields on
$M$ constitute a collection of infinitesimal local symmetries in the sense
of \cite{lw}. Hence, to each $\xi^a$ we may associate a Noether current
$(n-1)$-form, $\bf j$, defined by
\begin{equation}
{\bf j} = {\bf \Theta}(\phi, {\cal L}_{\xi} \phi) - \xi \cdot {\bf L}
\label{j}
\end{equation}
so that ${\bf j}$ is locally constructed out of the fields appearing in
${\bf L}$ and $\xi^a$. A standard calculation \cite{lw} shows that
\begin{equation}
d{\bf j} = - {\bf  E} { } {\cal L}_{\xi} \phi
\label{dj}
\end{equation}
so that ${\bf j}$ is closed whenever the equations of motion are satisfied.
Since $\bf j$ is closed for all $\xi^a$, it follows \cite{w} that there exists
an
$(n-2)$-form, $\bf Q$ -- locally constructed out of the fields appearing in
${\bf L}$ and $\xi^a$ -- such that when evaluated on solutions to the
equations of motion, we have,
\begin{equation}
{\bf j} = d{\bf Q}
\label{Q}
\end{equation}
Since ${\bf j}$ depends linearly on $\xi^a$, we adopt the explicit algorithm
provided by lemma 1 of
\cite{w} to uniquely define ${\bf Q}$, from which it follows that
${\bf Q}$ depends on no more than
$(k-1)$ derivatives of $\xi^a$, where
$k$ denotes the
highest derivative of any dynamical field occurring in ${\bf L}$.
(Note, however, that ${\bf Q}$ is unique up to addition of a closed -- and,
hence, exact \cite{w} -- $(n-2)$-form
locally constructed from the fields
appearing in ${\bf L}$ and from $\xi^a$, so the
integral of ${\bf Q}$ over any closed $(n-2)$-dimensional
surface, $\Sigma$, is uniquely defined by eq.(\ref{Q}) alone.)
We shall refer to ${\bf Q}$ as the
{\it Noether charge $(n-2)$-form} \cite{sim} relative to $\xi^a$, and its
integral over a closed surface, $\Sigma$, will be referred to as the
{\it Noether charge} of $\Sigma$ relative to $\xi^a$.

	The key identity upon which our derivation of the first law of
black hole mechanics will be based is obtained by considering the variation
of eq.(\ref{j}) resulting from an arbitrary variation, $\delta \phi$, of the
dynamical fields off of an arbitrary solution $\phi$. We have,
\begin{equation}
\delta {\bf j} = \delta [{\bf \Theta}(\phi, {\cal L}_{\xi} \phi)]
- \xi \cdot \delta {\bf L}
\label{delj1}
\end{equation}
(Note that $\xi^a$ is held fixed in this variation, i.e., we require that
$\delta \xi^a = 0$.)
However, by eq.(\ref{lag}), we have,
\begin{eqnarray}
\xi \cdot \delta {\bf L} & = & \xi \cdot [{\bf  E} \delta \phi
+ d {\bf \Theta}] \nonumber \\
 & = & {\cal L}_{\xi} {\bf \Theta} - d (\xi \cdot {\bf \Theta})
\label{lag2}
\end{eqnarray}
where the equations of motion, ${\bf E} = 0$,  for $\phi$ and
the identity (\ref{id}) were used in the second line.
Thus, we obtain,
\begin{equation}
\delta {\bf j} = \delta [{\bf \Theta}(\phi, {\cal L}_{\xi} \phi)]
- {\cal L}_{\xi} [{\bf \Theta}(\phi, \delta \phi)] + d (\xi \cdot {\bf \Theta})
\label{delj2}
\end{equation}
Note that in eq.(\ref{delj2}), no restrictions have been placed upon
$\delta \phi$ or $\xi^a$.

	Our next step is to identify certain ``surface terms" appearing
in eq.(\ref{delj2}). First, we require $\nabla_a$ to be
invariant under
the diffeomorphisms generated by $\xi^a$. This requirement
holds in the usual case where $\xi^a$ is taken to be a coordinate vector
field and $\nabla_a$ is taken to be the coordinate derivative operator of
that coordinate system; it also will hold
in our main application below where
$\nabla_a$ will be taken to be the derivative operator of the unperturbed
metric and $\xi^a$ is a Killing field of that metric.) In that case, the first
two terms on the right side of eq.(\ref{delj2}) combine to yield
\begin{equation}
\delta [{\bf \Theta}(\phi, {\cal L}_{\xi} \phi)]
- {\cal L}_{\xi} [{\bf \Theta}(\phi, \delta \phi)]
=  {\bf \Omega}(\phi, \delta \phi, {\cal L}_{\xi} \phi)
\label{sym2}
\end{equation}
and eq.(\ref{delj2}) becomes simply,
\begin{equation}
\delta {\bf j} =  {\bf \Omega}(\phi, \delta \phi, {\cal L}_{\xi} \phi)
+ d (\xi \cdot {\bf \Theta})
\label{delj3}
\end{equation}
When integrated over a Cauchy surface, ${\cal C}$
of the unperturbed solution,
eq.(\ref{delj3}) corresponds to eq.(3.22) of \cite{lw}, but eq.(\ref{delj3})
contains vital additional information concerning the ``surface term",
$d (\xi \cdot {\bf \Theta})$,
which did not appear
in \cite{lw}, since attention there was restricted to
the case of compact ${\cal C}$. Comparison of
eq.(\ref{delj3}) with
Hamilton's equations of motion shows that if a Hamiltonian, $H$,
corresponding to evolution by $\xi^a$ exists
on phase space, then $H$ must
satisfy,
\begin{equation}
\delta H =  \delta \int_{{\cal C}} {\bf j} -
\int_{{\cal C}} d (\xi \cdot {\bf \Theta})
\label{ham}
\end{equation}
where, in this equation, projection of the right side to phase space (in the
manner discussed in \cite{lw}) should be understood.
This shows that apart from the ``surface term" $d (\xi \cdot {\bf \Theta})$,
the Noether current, ${\bf j}$, acts as a
Hamiltonian density.

	We now further restrict attention to the case where $\delta \phi$
satisfies the linearized equations of motion, so that both $\phi$ and its
variation are solutions. Then we may replace ${\bf j}$ and its variation
by $d {\bf Q}$ in eqs.(\ref{delj3}) and (\ref{ham}).
It then can be seen immediately from eq.(\ref{ham}) that the Hamiltonian
-- if it exists -- is purely a ``surface term". In an asymptotically flat
spacetime, it is natural to associate the value of the surface contribution
to the Hamiltonian from infinity with the corresponding ``conserved
quantity" associated with $\xi^a$ in the manner of \cite{rt}. In other
words, {\it if} the theory admits a suitable definition of the ``canonical
energy", ${\cal E}$, associated with
an asymptotic time translation, $t^a$,
and of the ``canonical angular momentum"
${\cal J}$, associated with an asymptotic rotation, $\varphi^a$, the
variations of these quantities
should be given by the formulas
\begin{equation}
\delta {\cal E} = \int_{\infty} (\delta {\bf Q}[t] - t \cdot {\bf \Theta})
\label{E}
\end{equation}
\begin{equation}
\delta {\cal J} = - \int_{\infty} \delta {\bf Q}[\varphi]
\label{J}
\end{equation}
where the integrals are taken over an $(n-2)$-dimensional sphere
at infinity and the term $\varphi \cdot {\bf \Theta}$ does not appear
in eq.(\ref{J}) because $\varphi^a$ is assumed to be tangent to this
sphere. Thus, if one can find an $(n-1)$-form, ${\bf B}$, such that
\begin{equation}
\delta \int_{\infty} t \cdot {\bf B} = \int_{\infty} t \cdot {\bf \Theta}
\label{B}
\end{equation}
the canonical energy and angular momentum can be defined by,
\begin{equation}
{\cal E} = \int_{\infty} ({\bf Q}[t] - t \cdot {\bf B})
\label{E'}
\end{equation}
\begin{equation}
{\cal J} = - \int_{\infty} {\bf Q}[\varphi]
\label{J'}
\end{equation}
Note that ${\cal E}$  corresponds to the ``ADM mass" of
general relativity
plus possible additional contributions from any long-range matter fields
that may be present; see \cite{sw} for explicit discussion of the case
of the Yang-Mills field. Note also that for the Hilbert
Lagrangian of general relativity,
the expressions $\int_{\infty} {\bf Q}[t]$ and
$- \int_{\infty} {\bf Q}[\varphi]$ correspond -- up to numerical factors --
to the Komar expressions for mass and angular momentum. The presence
of the ``extra term" $t \cdot {\bf B}$ in eq.(\ref{E'}) accounts for why
different relative numerical factors must be chosen in the Komar formulas
for these quantities.) It
is, of course, a nontrivial condition on a theory that it admit a notion of
asymptotic flatness such that ${\cal E}$ and
${\cal J}$ are well defined. In the
following, I shall assume that this is the case, and derive the first law of
black hole mechanics for such a theory.

	Consider, now, a stationary black hole solution with a bifurcate Killing
horizon, with
bifurcation $(n-2)$-surface $\Sigma$. Choose $\xi^a$
to be the Killing field which vanishes on $\Sigma$, normalized so that
\begin{equation}
\xi^a = t^a + \Omega^{(\mu)}_H \varphi^a_{(\mu)}
\label{hkvf}
\end{equation}
where $t^a$ is the stationary Killing field (with unit norm at infinity)
and summation over ${\mu}$ is understood.
(This equation both picks out a particular family of axial Killing fields,
$\varphi^a_{(\mu)}$, acting in orthogonal planes,
and defines the ``angular velocity of the horizon",
$\Omega^{(\mu)}_H$.) Choose $\nabla_a$ to be the derivative operator of
this solution, so that $\nabla_a$ is invariant under the isometries
generated by $\xi^a$. Then eq.(\ref{sym2}) holds,
and, in addition, the right side now vanishes since
${\cal L}_{\xi} \phi = 0$. Let $\delta \phi$ be an arbitrary,
asymptotically flat solution of the linearized equations. Then, the
fundamental identity eq.(\ref{delj2}) yields simply
\begin{equation}
d (\delta{\bf Q})  = d (\xi \cdot {\bf \Theta})
\label{delQ}
\end{equation}
Choose ${\cal C}$ be an asymptotically flat hypersurface with
``interior boundary" $\Sigma$. Integrating eq.(\ref{delQ}) over ${\cal C}$,
taking into account eqs. (\ref{E}), (\ref{J}), and (\ref{hkvf})
together with the fact that $\xi^a$
vanishes on $\Sigma$, we obtain
\begin{equation}
\delta \int_{\Sigma} {\bf Q} = \delta {\cal E}
- \Omega^{(\mu)}_H \delta {\cal J}_{(\mu)}
\label{fl}
\end{equation}
Equation (\ref{fl}) corresponds precisely to the first law of black hole
mechanics as derived by Hamiltonian methods \cite{sw}.
However, eq.(\ref{fl})
has the advantage over this previous derivation
that the surface term arising from the black hole
has now been explicitly identified
as the variation of the Noether charge of $\Sigma$.

	Equation (\ref{fl}) still is not of the desired form in the sense that
the left side of eq.(\ref{fl}) has not yet been written as $\kappa$ times the
variation of a local, geometrical quantity on $\Sigma$,
since ${\bf Q}$
is locally constructed from $\xi^a$ and its derivatives as well as from
the fields appearing in the Lagrangian.
However, we now will show that the desired form of the first law
holds when we further
restrict attention to the case where $\delta \phi$ describes a perturbation
to a nearby stationary black hole. First, we note that
any derivative, $\nabla_{a_1} ... \nabla_{a_n} \xi^b$ of any Killing field
$\xi^a$ can be re-expressed in terms of a linear combination of $\xi^a$
and its first derivative, $\nabla_{a} \xi_{b}$, with
coefficients depending upon the Riemann curvature and its derivatives
(see, e.g., eq.(C.3.6) of \cite{wa}). Next, we note that on $\Sigma$ we have
$\xi^a = 0$ and $\nabla_{a} \xi_{b} = \kappa \epsilon_{ab}$, where
$\epsilon_{ab}$ denotes the bi-normal to $\Sigma$.
Now define the $(n-2)$-form
$\tilde{{\bf Q}}$ on $\Sigma$ by the following algorithm:
Express ${\bf Q}$ in terms of $\xi^a$ and $\nabla_{a} \xi_{b}$
by eliminating the higher derivatives of $\xi^a$, as described above.
Then set $\xi^a = 0$ and replace $\nabla_{a} \xi_{b}$ by
$\epsilon_{ab}$. Since any reference to $\xi^a$ has been
eliminated, we see that $\tilde{{\bf Q}}$
is locally constructed out of the fields appearing
in ${\bf L}$. Furthermore, since $\tilde{{\bf Q}}$ on $\Sigma$ is
determined by a well defined
algorithm whose only input is a Lagrangian ${\bf L}$ which is
invariant under diffeomorphisms of the dynamical fields, $\phi$,
(see eq.(\ref{dif}) above) it follows that
$\tilde{{\bf Q}}$ is similarly invariant under spacetime diffeomorphisms
which map $\Sigma$ into itself. Thus,
$\tilde{{\bf Q}}$ is a ``local, geometrical quantity" on $\Sigma$. Finally, it
is
worth noting that $\tilde{{\bf Q}}$ is just the Noether
charge $(n-2)$- form associated with the Killing field $\tilde{\xi}^a
= \kappa \xi^a$, i.e., $\tilde{\xi}^a$ is the horizon Killing field
normalized so as to have unit surface gravity.

	Now identify the unperturbed and perturbed
stationary black hole spacetimes in such a way that the Killing
horizons of the two spacetimes coincide, and the unit surface gravity
horizon Killing fields $\tilde{\xi}^a$ coincide in a neighborhood of the
horizons. (That this always can be done follows from the general formula
for Kruskal-type coordinates given in \cite{rw}. Note, however, that for a
perturbation which changes the surface
gravity, we cannot identify the two
spacetimes so that the two horizon Killing fields
coincide on the horizon when normalized via eq.(\ref{hkvf});
in addition, since we take $\delta t^a = \delta \varphi^a_{(\mu)} = 0$ near
infinity, for a
perturbation which changes $\Omega^{(\mu)}_H$, the requirement that
$\delta \xi^a = 0$
precludes us from choosing $\xi^a$ even to be proportional to the horizon
Killing field near infinity in the perturbed spacetime.) Then, on $\Sigma$
we have,
\begin{equation}
\delta {\bf Q} = \kappa \delta \tilde{{\bf Q}}
\label{Q'}
\end{equation}
where $\kappa$ is the surface gravity of the unperturbed black hole.
Hence, for perturbations to nearby stationary black holes, the first law
of black hole mechanics (\ref{fl}) takes the form
\begin{equation}
\frac{\kappa}{2 \pi} \delta S = \delta {\cal E} -
\Omega^{(\mu)}_H \delta {\cal J}_{(\mu)}
\label{fl'}
\end{equation}
where the ``black hole entropy", $S$, is defined by
\begin{equation}
S = 2 \pi \int_{\Sigma} \tilde{{\bf Q}}
\label{S}
\end{equation}
Thus, we have established the existence and
``local, geometrical character" of the notion of
black hole entropy,
$S$, in a general theory of gravity \cite{siw}.

	Our ``local, geometrical" formula (\ref{S}) for the entropy of a
stationary black hole suggests the following generalization to the
non-stationary case:
For an arbitrary cross-section, ${\Sigma}'$, of the horizon of a non-stationary
black hole, construct
$\tilde{{\bf Q}}$ by exactly the same mathematical algorithm as used
above for the bifurcation surface, $\Sigma$, of a Killing horizon.
Then $2 \pi$ times the integral
of $\tilde{{\bf Q}}$ over ${\Sigma}'$ yields a candidate expression for
the black hole entropy at ``time" ${\Sigma}'$. The viability of this
proposed definition is presently under investigation \cite{iw}.

	Note that eq.(\ref{fl'}) has been derived
only for the case of perturbations
to nearby stationary black holes, even though eq.(\ref{fl}) holds in the
more general case of
non-stationary perturbations. However, since $\delta \xi^a = 0$ and,
on $\Sigma$, we have $\xi^a = 0$, it follows that
$\delta [\nabla_b \xi^a ] = 0$ on $\Sigma$. From this, it follows that for
non-stationary perturbations, we have, on $\Sigma$,
\begin{equation}
\delta [\nabla_{[a} \xi_{b]} ] = \kappa \delta \epsilon_{ab} + w_{ab}
\label{lg}
\end{equation}
where, as before, $\epsilon_{ab}$ denotes the binormal to $\Sigma$,
and $w_{ab}$ is purely ``normal-tangential", i.e., it vanishes when both
of its indices are projected into $\Sigma$ or both projected normal to
$\Sigma$. It then follows from the existence of a reflection isometry about
$\Sigma$ (see lemma 2.3 of \cite{kw}) that the $w_{ab}$-term makes
no contribution to the variation of ${\bf Q}$. It then can be seen that
for sufficiently ``low derivative" theories
where ${\bf Q}$ depends only upon $\xi^a$
and its first antisymmetrized derivative (as occurs, in particular, in general
relativity and in $(1+1)$-dimensional theories), eq.(\ref{Q'})
continues to hold for non-stationary perturbations. Thus, in such theories,
the first law of black hole mechanics continues to hold in the form
(\ref{fl'}), with $S$ given by (\ref{S}). The nature of the first law for
non-stationary perturbations in more general theories is presently under
investigation \cite{iw}.

	The fact that for stationary black holes, $S$ is just $2 \pi$ times the
Noether charge of the horizon Killing field (normalized to have unit
surface gravity) implies that for an initially stationary black hole which
undergoes a dynamical process and later ``settles down" to a stationary
final state, the net change in black hole entropy is just the
total flux through the horizon of Noether current associated with a
suitable ``time translation" on the horizon. This suggests a possible
relationship between the validity of the second law of black hole
mechanics in a theory and positive energy properties of that theory. It
also suggests some possible approaches toward establishing (or
disproving) the second law in general theories of gravity. These issues
also are under investigation \cite{iw}.

	Finally, we consider the relationship between the results of
this paper and the formula
for black hole entropy obtained via the ``Euclidean approach" in the
manner first given in \cite{gh}. We begin by noting that since
${\cal L}_{\xi} \phi = 0$, the Noether current (\ref{j}) associated with the
horizon Killing field, $\xi^a$, of a stationary black
hole is simply
\begin{equation}
{\bf j} = - \xi \cdot {\bf L}
\label{j'}
\end{equation}
Let ${\cal C}$ be an asymptotically
flat hypersurface with ``interior boundary" $\Sigma$. Integrating
eq.(\ref{j'}) over ${\cal C}$ and taking into account eqs.(\ref{Q}),
(\ref{hkvf}), (\ref{E'}), and (\ref{J'}), we obtain,
\begin{equation}
 {\cal E} - \Omega^{(\mu)}_H {\cal J}_{(\mu)} - \int_{\Sigma} {\bf Q}
= - \int_{{\cal C}}\xi \cdot {\bf L} - \int_{\infty} t \cdot {\bf B}
\label{action}
\end{equation}
Now, the ``Euclidean action", $I$, corresponds to
\begin{equation}
I = - \frac{2 \pi}{\kappa} [\int_{{\cal C}}\xi \cdot {\bf L}
+ \int_{\infty} t \cdot {\bf B}]
\label{I}
\end{equation}
More precisely, in the static case, the right side of eq.(\ref{I}) equals
what would be obtained by integrating the suitably analytically
continued Lagrangian, ${\bf L}' = {\bf L} + d {\bf B}$, over a ``Euclidean
section", constructed by replacing the Killing parameter,
$t$, by $\tau = i t$,
and then periodically identifying $\tau$ with period ${2 \pi}/ \kappa$
(see, e.g., \cite{ha} for further details). In the
stationary but non-static case, there is no such thing as a
``Euclidean section", but the right side of eq.(\ref{I}) corresponds to what
researchers mean by the ``Euclidean action" in that case. Thus, we obtain
the following formula for $I$,
\begin{equation}
\frac{\kappa}{2 \pi} I =
{\cal E} - \Omega^{(\mu)}_H {\cal J}_{(\mu)} - \int_{\Sigma} {\bf Q}
\label{I'}
\end{equation}
Now, in the Euclidean approach, $\frac{\kappa}{2 \pi} I$ is
identified as the thermodynamic potential
of the black hole \cite{gh}. This leads
immediately to the the following formula for black hole entropy,
\begin{eqnarray}
 \frac{\kappa}{2 \pi} S & = & {\cal E} - \Omega^{(\mu)}_H {\cal J}_{(\mu)}
- \frac{\kappa}{2 \pi} I \nonumber \\
& = & \int_{\Sigma} {\bf Q}
\label{S'}
\end{eqnarray}
which agrees with eq.(\ref{S}). Thus, we have shown that the ``Euclidean
procedure" for obtaining black hole entropy gives the same result as
obtained by our method.

	This research was supported in part by
National Science Foundation
grant PHY-9220644 to the University of Chicago.


\begin{thebibliography}{99}
\bibitem{haw} S.W. Hawking,  Commun. Math. Phys.
{\bf 43}, 199 (1975).
\bibitem{bch}
J.M.  Bardeen, B. Carter, S.W. Hawking, Commun.
Math. Phys. {\bf 31}, 161 (1973).
\bibitem{sw}
D. Sudarsky and R.M. Wald, Phys. Rev. D {\bf 46}, 1453 (1992);
R.M. Wald, `` The first law of black hole mechanics" in
{\it Directions in General Relativity}, vol. 1, ed. by B.L. Hu, M. Ryan, and
C.V. Vishveshwara (Cambridge University Press, Cambridge, 1993).
See also J.D. Brown and J.W. York Phys. Rev. {\bf D47},
1407 and 1420 (1993).
\bibitem{fro} V.P. Frolov, Phys. Rev. {\bf D46}, 5383 (1992).
\bibitem{jm} T.A. Jacobson and R.C. Myers, Phys. Rev. Lett.
{\bf 70}, 3684 (1993)
\bibitem{jac} I wish to thank Ted Jacobson for bringing this deficiency
to my attention.
\bibitem{gh} G.W. Gibbons and S.W. Hawking, Phys. Rev. {\bf D15},
2752 (1977).
\bibitem{iw} V. Iyer and R.M. Wald, to be published.
\bibitem{he} S. W. Hawking and G. F. R. Ellis,
{\it The Large Scale Structure of Space-Time} (Cambridge
University Press, Cambridge, 1973).
\bibitem{wa} R. M. Wald, {\it General Relativity}
(University of Chicago Press, Chicago, 1984).
\bibitem{zl} It is proven in \cite{rw} that in an arbitrary theory of gravity,
global nonsingularity of the
event horizon of a stationary black hole implies that the zeroth law holds.
However, there are no compelling physical grounds to impose such a
global nonsingularity condition, since for a black hole formed by the
collapse of matter, the parallelly propagated curvature singularity
found in \cite{rw} would be ``covered up" by the collapsing matter.
\bibitem{rw} I. Racz and R.M. Wald, Class and Quant. Grav.,
{\bf 9}, 2643 (1992).
\bibitem{kw} B.S. Kay and R.M. Wald, Phys. Rep. {\bf 207},
49 (1991).
\bibitem{lw} J. Lee and R.M. Wald, J. Math. Phys. {\bf 31},
725 (1990).
\bibitem{w} R.M. Wald, J. Math. Phys. {\bf 31},
2378 (1990). The results of this reference also can be derived using
the ``free variational bicomplex"; see I.M. Anderson in
{\it Mathematical Aspects of Classical Field Theory}, ed. by M. Gotay,
J. Marsden, and V. Moncrief, Cont. Math. {\bf 132}, 51 (1992).
\bibitem{sim} This quantity corresponds
to the ``gravitational field strength"
discussed in W. Simon, Gen. Rel. and Grav. {\bf 17}, 439 (1985).
\bibitem{rt} T. Regge and C. Teitelboim, Ann. Phys. {\bf 88}, 286 (1974).
\bibitem{siw} A much more indirect argument for this conclusion has been
given independently by J. Simon and B. Whiting (to be published), based
upon the Euclidean approach. See also, M. Visser (to be published).
\bibitem{ha} S.W. Hawking, ``The Path Integral Approach to Quantum
Gravity", in {\it General Relativity, an Einstein
Centennary Survey}, ed. by S.W. Hawking and W. Israel
(Cambridge University Press, Cambridge, 1979)

\end{thebibliography}
\end{document}